\documentclass[%
 reprint, 
superscriptaddress,
 amsmath,amssymb,
 aps,
pra, 
longbibliography, floatfix ]{revtex4-1}

\usepackage{natbib} 
\usepackage{amsmath} 
\usepackage{graphicx} 
\usepackage{longtable}
\usepackage{multirow}
\usepackage{amsmath}
\usepackage{epstopdf}
\usepackage{placeins}

\usepackage{xcolor}  

\usepackage[english]{babel}

\makeatletter
\def\bbl@set@language#1{%
  \edef\languagename{%
    \ifnum\escapechar=\expandafter`\string#1\@empty
    \else\string#1\@empty\fi}%
  \@ifundefined{babel@language@alias@\languagename}{}{%
    \edef\languagename{\@nameuse{babel@language@alias@\languagename}}%
  }%
  \select@language{\languagename}%
  \expandafter\ifx\csname date\languagename\endcsname\relax\else
    \if@filesw
      \protected@write\@auxout{}{\string\select@language{\languagename}}%
      \bbl@for\bbl@tempa\BabelContentsFiles{%
        \addtocontents{\bbl@tempa}{\xstring\select@language{\languagename}}}%
      \bbl@usehooks{write}{}%
    \fi
  \fi}
\newcommand{\DeclareLanguageAlias}[2]{%
  \global\@namedef{babel@language@alias@#1}{#2}%
} \makeatother

\DeclareLanguageAlias{en}{english}

\begin{document}

\title{Exploring the Structure of Misconceptions in the Force Concept Inventory with Modified Module Analysis}
\author{James Wells}
\affiliation{%
    W.~M.~Keck Science Department of Claremont McKenna, Pitzer, and Scripps Colleges, Claremont CA, 91711
}%
\author{Rachel Henderson}
\affiliation{%
    Michigan State University, Department of Physics and Astronomy,
    East Lansing MI, 48824
}%
\author{John Stewart}
\email{jcstewart1@mail.wvu.edu}
\affiliation{%
    West Virginia University, Department of Physics and Astronomy,
    Morgantown WV, 26506
}%
\author{Gay Stewart}
\affiliation{%
    West Virginia University, Department of Physics and Astronomy,
    Morgantown WV, 26506
}%
\author{Jie Yang}
\affiliation{%
    West Virginia University, Department of Physics and Astronomy,
    Morgantown WV, 26506
}%
\author{Adrienne Traxler}
\affiliation{%
    Wright State University, Department of Physics,
    Dayton OH, 45435
}%

\date{\today}

\begin{abstract}
Module Analysis for Multiple-Choice Responses (MAMCR) was applied
to a large sample of Force Concept Inventory (FCI) pretest and
post-test responses ($N_{pre}=4509$ and $N_{post}=4716$) to
replicate the results of the original MAMCR study and to
understand the origins of the gender differences reported in a
previous study of this data set. When the results of MAMCR could
not be replicated, a modification of the method was introduced,
Modified Module Analysis (MMA). MMA was productive in
understanding the structure of the incorrect answers in the FCI,
identifying 9 groups of incorrect answers on the pretest and 11
groups on the post-test. These groups, in most cases, could be
mapped on to common misconceptions used by the authors of the FCI
to create distactors for the instrument.  Of these incorrect
answer groups, 6 of the pretest groups and 8 of the post-test
groups were the same for men and women. Two of the male-only
pretest groups disappeared with instruction while the third
male-only pretest group was identified for both men and women
post-instruction. Three of the groups identified for both men and
women on the post-test were not present for either on the pretest.
The rest of the identified incorrect answer groups did not
represent misconceptions, but were rather related to the the
blocked structure of some FCI items where multiple items are
related to a common stem. The groups identified had little
relation to the gender unfair items previously identified for this
data set, and therefore, differences in the structure of student
misconceptions between men and women cannot explain the gender
differences reported for the FCI.

\end{abstract}

\maketitle

\section{Introduction}
The ``gender gap,'' gender differences between the scores of men
and women on the Force Concept Inventory (FCI) \cite{hestenes1992}
and other instruments developed by Physics Education Research
(PER), has been extensively studied \cite{madsen2013}. For the
FCI, a substantial number of studies have suggested that some of
the gender differences observed resulted from different response
patterns of men and women to a subset of the items in the
instrument \cite{traxler2018}. The origin of these differential
response patterns is, however, unknown. The purpose of this study
is to apply Module Analysis for Multiple-Choice Responses
introduced by Brewe {\it et al.} \cite{brewe2016} to a large
sample of FCI responses known to contain a subset of items which
produce substantially different response patterns for men and
women in order to determine if the structure of the misconceptions
of men and women differ on these items.

\subsection{Background Studies}

This work will draw heavily from three previous studies which will
be referenced as Study 1, Study 2, and Study 3 in this work.

\subsubsection{Study 1: Module Analysis \label{Study1}}
Study 1 introduced Module Analysis for Multiple Choice Responses
(MAMCR) to analyze concept inventory data at the level of
individual responses to the items \cite{brewe2016}. Unlike many
analysis techniques applied to FCI data, which consider only a
student's overall score or only the correct answers to individual
items, MAMCR considers each answer choice a student selected in
order to provide a fine-grained examination of students'
misconceptions of Newtonian physics and to allow instructors to
target specific errors.

MAMCR is based on network analytic techniques \cite{newman,
zweig}. A network is represented by a graph where nodes  are
connected to one another by edges. Edges can be weighted, where the
value of the weight represents some aspect of the interaction.
Network analysis is a highly successful and versatile
set  of methods which have been applied to a variety of problems
including the probability of homicide victimization among people
living in a disadvantaged neighborhood \cite{papachristos}, the
mapping of functional networks in the brain from electrical
signals \cite{devico}, passing patterns of soccer teams in the World Cup \cite{pena},
and the response of plants to bacterial infection \cite{zheng}.

Study 1 examined the FCI post-test scores of 143 first-year
physics majors at a Danish university. The sample was $78\%$ male
and scored relatively highly on the exam: pre-test $65\pm22\%$ and
post-test $81\pm18\%$.

To analyze the FCI, each response was assigned to a node in the
network; for example, if a student selected the choice ``D'' on
FCI item 1, then 1D would be a node. An edge was added for each
time a student selected two responses; for example, if a student
selected 1D and 2E, then an edge was drawn connecting 1D and 2E.
The correct responses were removed from the network leaving the
network of incorrect responses.

In order to find connected responses in the network, a Community
Detection Algorithm (CDA) was applied to the network.  There are
many different types of CDAs \cite{fortunato}; in Study 1, the
Infomap algorithm was chosen \cite{rosvall}.

Study 1 identified nine modules, each representing a separate
misconception in student thinking. Two modules could be clearly
interpreted: module 1 ``the impetus model'' and module 2 ``more
force yields more results.'' The other seven modules were more
difficult to interpret.

In Study 1, Brewe \emph{et al.} emphasized that the results of
this study should be generalized with care. The group of students
tested was small, unusually high scoring, and had limited
diversity. Likewise, there were several choices made during the
process of applying MAMCR to the data which could have been made
differently. Both the choice of sparsification method (described
in Sec.~\ref{sec:ModuleAn}) and the decision of how to group
responses that cluster together only on some of the one-thousand
applications of Infomap was somewhat arbitrary, as was the
interpretation of the meaning of the modules. As will be seen in
Sec.~\ref{sec:ModuleAn}, our data set required different choices
to be made.

\subsubsection{Study 2: Item Fairness and the FCI}
\label{sec:study2} In Study 2, Traxler {\it et al.}
\cite{traxler2018} explored item-level gender fairness of the FCI
using Classical Test Theory \cite{crocker_introduction_1986}, Item
Response Theory \cite{de2013theory}, and Differential Item
Functioning (DIF) \cite{holland1985alternate,holland1988}
analysis. An item is fair to men and women if men and women of
equal overall ability score equally on the item. Using three
samples, a graphical analysis identified five FCI items that were
substantially unfair to women: item 14 (bowling ball falling out
of an airplane), items 21 through 23 (sideways-drifting rocket
with engine turning on and off), and item 27 (a large box being
pushed across a horizontal floor). A further DIF analysis, which
controlled for the student's overall post-test score, identified
eight items on the FCI as substantially unfair. Two of these were
unfair to men: item 9 (speed of a puck after it receives a kick)
and item 15 (a small car pushing a large truck). These eight items
included the five items identified in the graphical analysis along
with items 9, 12 (the trajectory of a cannon ball shot off of a
cliff), and 15. Many of the unfair items had been identified as
unfair in previous studies. Overall, Study 2 demonstrated that
eliminating all unfair items on the FCI to create a fair
instrument reduced the gender gap by 50\% in the largest sample.

This work, however, could not identify the source of the
unfairness.  The distribution of student responses was analyzed.
Focusing on the five items that were identified with the graphical
analysis and the DIF analysis, incorrect female responses were
predominately one of the distractors in each of the FCI items;
however, the distractors chosen by the male students were less
uniform in all five FCI items. Overall, Study 2 concluded that no
physical principle or common misconception could explain the
unfairness identified FCI items; however, this conclusion was
drawn from a qualitative inspection of the items. The current
study builds on the work in Study 2 by performing a quantitative
analysis of the incorrect responses of men and women.

\subsubsection{Study 3: Multidimensional Item Response Theory and the FCI}

The study in the current work applies network analytic methods to
understand the incorrect answer structure of the FCI. This
structure might be influenced by features of the FCI which produce
correlations between the correct answers. If a consistent
misconception is being applied, it would form an alternate
incorrect answer to sets of related correct answers. Study 3
examined the correct answer structure of the FCI using both
exploratory and confirmatory methods \cite{stewart2018}.
Exploratory factor analysis (EFA) suggested that the practice of
``blocking'' items produced correlations between the items within
the block. A block of items is a sequence of items which all refer
to a common stem or where one item refers to a previous item. The
FCI contains item blocks \{5, 6\}, \{8, 9, 10, 11\}, \{15, 16\},
\{21, 22, 23, 24\}, and \{25, 26, 27\}. Study 3 reported that
often the factors identified by EFA strongly loaded on items in
the same block, suggesting that blocking was generating
correlations among the items in the block. Study 3 went on to
produce a detailed model of the reasoning required to solve the
FCI. Multidimensional Item Response Theory (MIRT) was used to test
alternate models and allowed the identification of an optimal
model. This model allowed the identification of groups of items
with very similar solution structure: \{5, 18\}, \{6, 7\}, \{17,
25\}, and \{4, 15, 28\}. Study 3 only included the first item in a
block in the analysis and it is likely that item 16 should be
added to the last block which represents Newton's 3rd law items.
This mapping of item blocks and groups with similar solution will
be important to understanding the incorrect answer structure
presented in this work.

\subsection{Previous Studies of the FCI}
The FCI, either in aggregate or disaggregating by gender, is one
of the most studied instruments in PER. The present study examines
item-level structure disaggregated by gender. The structure of the
incorrect answers is examined to identify coherent patterns of
incorrect answers.

\subsubsection{Exploratory Analyses of the FCI}

Many studies have examined the structure of the FCI, primarily
using EFA. These studies began soon after the publication of the
FCI when Huffman and Heller \cite{huffman_what_1995} failed to
extract the factor structure suggested by the authors of the
instrument \cite{hestenes1992}. For a sample of 145 high school
students, Huffman and Heller found only two post-test factors:
``Newton's 3rd law'' and ``Kinds of Forces.'' The small number of
factors may have resulted from a very conservative factor
selection criteria. In the same study, for a sample of 750
university students, only one factor was identified: ``Kinds of
Forces.'' A later work by Scott, Schumayer and Gray
\cite{scott2012exploratory} applied EFA to the FCI post-test
scores of a sample of 2150 students and found an optimal model
with 5 factors; however one of the factors explained much of the
variance. The result that a single factor explains the majority of
the variance is fairly robust and is further supported by the high
Cronbach alpha values reported in Study 2 and by Lasry {\it et
al.} \cite{lasry2011puzzling}.  Scott and Schumayer
\cite{scott2015} replicated their 5 factor analysis using MIRT on
the same sample. Semak {\it et al.} \cite{semak2017} reported
optimal models with 5 factors on the pretest and 6 factors on the
post-test when exploring the evolution of student thinking for 427
algebra- and calculus-based introductory physics students. Study 3
also performed EFA using MIRT and reported 9 factors as optimal.

\subsubsection{Gender and the FCI}

In an extensive review of gender differences on physics concept
inventories \cite{madsen2013},  men outperformed women by $13\%$
on pretests and $12\%$ on post-tests of conceptual mechanics: the
FCI and the Force and Motion Conceptual Evaluation
\cite{thornton1998}.

Many reasons have been explored to explain these differences.
Differences in high school physics class election
\cite{edu2009,nces2015,ets2016} may cause differences in college
physics grades \cite{sadler2001,hazari2007}. In addition,
many studies have identified gender differences in academic course grades with women
generally outperforming men \cite{voyer2014gender}; these differences may influence
conceptual inventory performance. Cognitive
differences have also been advanced as explanations of academic
gender differences
\cite{maeda2013,halpern,hyde1988gender,hyde1990gender} with women
scoring generally higher on verbal reasoning tasks and men scoring generally higher on
spatial reasoning tasks; however, cognitive differences between men and women are
fine grained with differences within the subskills of a single discipline \cite{ets1997}. Psychocultural factors have also been
advanced as explanations of academic performance differences
including mathematics anxiety \cite{else2010cross,ma1999meta},
science anxiety \cite{mallow1982,udo2004,mallow2010}, and
stereotype threat \cite{shapiro2012}. For a more detailed
discussion about the many sources that may influence the overall
gender differences on physics conceptual inventories, see
Henderson \textit{et al.} \cite{henderson2017}.

\subsubsection{Item Fairness and the FCI}

In addition to student-centered explanations for conceptual
inventory gender differences, bias in the individual FCI items has
been investigated as a source of these gender differences.
McCullough and Meltzer \cite{mccullough_differences_2001} randomly
gave students the original FCI or a version where each problem's
context was modified to be more stereotypically familiar to women.
In a sample of 222 algebra-based physics students, they found
significant differences in performance on items 14, 22, 23, and
29. In 2004, in a sample of non-physics students, McCullough used
a similar methodology \cite{mccullough_gender_2004}, and found
that female performance did not change while male performance
decreased on the modified contexts. Multiple studies have reported
item unfairness in unmodified items in the FCI
\cite{popp2011,dietz_gender_2012}.

Study 2 provides a thorough summary of research into the item
fairness of the FCI \cite{traxler2018}. Recent research has
suggested that other commonly used conceptual physics instruments
do not contain a substantial number of unfair items
\cite{henderson2018}.

\subsection{Misconception Research}

Since the early 1980's, student difficulties, most commonly known
as ``misconceptions'' or ``alternate conceptions/hypotheses,''
have been extensively studied within physics classrooms. The early
work done by Clement and colleagues
\cite{clement1982,clement1989,clement1993} qualitatively analyzing
the ``alternate view of the relationship between force and
acceleration'' that are grounded in students' experiences has
influenced much of the research examining conceptual understanding
in physics. Halloun and Hestenes
\cite{halloun1985initial,halloun1985common} further explored this
idea by collecting a taxonomy of ``common sense concepts'' that
conflict with student understanding of Newtonian mechanics.
Hestenes, Wells, and Swackhamer developed the FCI
\cite{hestenes1992} with the intent of measuring student
conceptual understanding of Newtonian theory, specifically
analyzing student misconceptions pre- and post-instruction
\cite{hake1998}.

\subsubsection{Misconceptions and the FCI}

The authors of the FCI provided a detailed description of the
misconceptions measured by the instrument \cite{hestenes1992}. A
summary of those misconceptions follow.

\vspace{6pt}

\noindent{\textit{Impetus.}} Dating back to pre-Galilean times,
the impetus model involves the idea that an object has a ``motive
power'' that can explain why an object remains in motion
regardless of any external forces
\cite{halloun1985common,hestenes1992}. Students with this
misconception do not fully understand Newton's 1st law. For
example, FCI items 6 and 7 describe a ball moving in a circle and
ask about the path the ball will take after it exits a circular
path. Selecting the circular trajectory after exiting the track
demonstrates the misconception that the ball has a circular
impetus.

\vspace{6pt}

\noindent{\textit{Active Force.}} The misconception that motion
implies force involves the idea that an object in motion must be
experiencing a force. This misconception involves a naive
understanding of the difference between velocity and acceleration
\cite{clement1982, hestenes1992} and demonstrates that Newton's
2nd law is not well understood. For example, items 5 and 18
describe an object moving in a circular path and ask about the
forces acting on the object. The motion implies force
misconception would predict that there is force in the direction
of the motion.
\vspace{6pt}

\noindent{\textit{Action/Reaction Pairs.}} The misconception that
the larger object exerts a greater force on a smaller object stems
from the ``dominance principle''
\cite{hestenes1992,halloun1985common}. This misconception
demonstrates that Newton's 3rd law is not well understood. For
example, items 4 and 15 describe a small car pushing a large truck
and ask to describe the forces between the two objects. The
``dominance principle'' misconception would predict the truck
exerts a larger force on the car than the car exerts on the truck.

\vspace{6pt}

\noindent{\textit{Concatenation of Influences.}} This
misconception involves the idea that forces influence with ``one
force winning out over the other'' \cite{hestenes1992}. This
misconception demonstrates that the superposition principle for
Newtonian forces is not well understood. For example, items 8 and
9 describe a hockey puck sliding horizontally at a constant speed
on a frictionless surface. These items ask for the path that the
hockey puck would take and the speed of the puck after it receives
a swift kick. The misconception of ``one force winning'' would
predict that the last force (i.e. the swift kick) determines the
motion and speed of the puck.

\vspace{6pt}

\noindent{\textit{Gravity.}} The misconception that gravity is not
a force stems from the Aristotelian physics idea that heavier
objects tend to move toward the center of the earth and lighter
objects tend to move away from the center of the earth
\cite{halloun1985common,hestenes1992}. For example, FCI items 1
and 2 describe two metal balls of different weights that are (1)
dropped at the same time and (2) rolled off of a horizontal table
at the same speed; the items ask about the amount of time it takes
for the two balls to hit the ground and the horizontal distance
traveled, respectively. The gravity misconception predicts that
the heavier ball falls faster and travels further.

Recently, quantitative studies have been used to begin to further
understand the misconception structure of the FCI. Scott and
Schumayer \cite{scott2017conceptual} applied EFA to all 150
responses, 5 per item, on the FCI pretest. The two most important
factors each contained responses from the majority of the items in
the FCI; contained both incorrect and correct responses; and mixed
conceptually very different correct reasoning as characterized by
the model in Study 3. For example, factor 1 contained correct
responses to questions on Newton's 1st law, Newton's 3rd law, one
and two-dimensional motion under gravity, and one and
two-dimensional motion ignoring gravity. Three of the six factors
showed evidence of students answering in patterns in the data set
(always selecting response ``A,'' ``C,'' or ``E'' when unsure of
the answer). Eaton, Vavruska, and Willoughby \cite{eaton2019}
replicated this work for both pretest and post-test data; no
consistent theme could be identified for multiple factors in their
study. The failure of these studies to identify an intelligible
factor structure containing items requiring related Newtonian
reasoning may indicate that factoring the incorrect and correct
items together in the same analysis is not productive or may be
seen as further support for Study 3 which concluded that EFA was
not a productive method to explore the FCI.

Scott and Schumayer provided additional analysis of two of their
factors using network analytic techniques \cite{scott2018central}.
As in this work, the network was constructed using the correlation
matrix; however, only correlations within the factors identified
in their early factor analysis were considered. This work reported
node centrality measures, but did not use the community detection
methods of Study 1.

\subsubsection{Misconception Research}

Many researchers have investigated students' conceptual
understanding exploring the misconceptions outlined above. Early
research  explored the overall common difficulties and beliefs
that students had about Newtonian mechanics
\cite{viennot1979,trowbridge1981,caramazza1981,peters1982,mccloskey1983,gunstone1987,camp1994}.
More recently, researchers have designed systematic studies to
explore student understanding and the epistemological development
of Newton's Laws of Motion
\cite{mcdermott1997,thornton1998,rosenblatt2011,erceg2014,waldrip2014}.
For example, Rosenblatt and Heckler developed a new assessment to
investigate student understanding of the relationship between
force, velocity, and acceleration \cite{rosenblatt2011}. This
study found that understanding the relationship between velocity
and acceleration was necessary to understanding the relationship
between velocity and force; however, the reverse was not
necessarily true.

In general, a misconception about mechanics can be defined as a
non-Newtonian reasoning principles directly related to the
physical systems addressed by Newtonian mechanics. Modeling
coherent patterns of student wrong answers as misconceptions is
only one of many ways to explain patterns of reasoning about
mechanics. Other important theories include knowledge in pieces
\cite{disessa1988,disessa1993,disessa1998} and ontological
categories \cite{chi1993,chi1994,slotta1995}. The
knowledge-in-pieces framework posits that student knowledge is
formed of a number of granular facts, p-prims, that are activated
either individually or collectively to produce a solution. In
general, both p-prims and misconceptions are small segments of
reasoning; p-prims are more general, while misconceptions are more
specific to the physical context. The ontological categories
framework is substantially different than either the misconception
view or the knowledge-in-pieces view; ontological categories
posits that incorrect student answers result from a
misclassification of a concept. For example, misclassification of
the concept of force as a substance that can be used up. A
substantial amount of research has also investigated how students'
conceptual knowledge changes over time \cite{duit2003}. In PER, Hammer proposed an
extension and unification  of the knowledge-in-pieces and misconception views
modeling both misconceptions and p-prims as ``resources'' \cite{hammer1996misconceptions,hammer_more_1996,hammer2000student}.
A resource was developed in analogy to a segment of computer code of
arbitrary complexity. A p-prim would represent a fundamental subset of the
code, while a misconception would represent a consistent misapplication
of the code. Unlike the misconception view, the resource view identifies
positive intellectual components which a instructor can activate to encourage
the knowledge construction process. The quantitative method in the present work
identified small segments of incorrect reasoning, and as such, cannot inform the
resource view of student knowledge.
In addition, the FCI was strongly developed within the misconception view and, therefore,
this work will as focus on that view.

The framework chosen, knowledge-in-pieces, misconceptions, or
ontological categories, has different consequences for instruction
or curriculum design in how they draw out and make use of student
ideas \citep{hammer_more_1996}. However, it is less clear that
this difference is measured by conceptual inventories. Incoherence
in student answers for the same concept might suggest a
knowledge-in-pieces view, where different problem contexts can
trigger different p-prims even if a physicist would see the
scenarios as isomorphic. However, the FCI was not designed to
measure this effect, and as such, a separate instrument designed
around the knowledge-in-pieces or ontological categories
frameworks is likely required to fully explore either framework.

In the text, we will primarily use the naive theory or
misconceptions framework \cite{etkina2005,bransford2000} with
notes in the Results where alternative frameworks  seem relevant.
Ultimately, while we call groups of incorrect answers
identified by network analytic techniques ``misconceptions,'' this
work is purely quantitative and cannot distinguish between the
various theoretical frameworks developed to explain incorrect
answering patterns.

\subsection{Research Questions}

This study attempted to reproduce the results of Study 1
disaggregated by gender. When the results were not reproduced, the
reasons for the failure of MAMCR were explored and a modification
to the algorithm proposed called ``Modified Module Analysis.'' The
modified algorithm was used to explore gendered differences in the
patterns of incorrect answers on the FCI.

In general, network analysis uses the term ``community'' and
``module'' interchangeably to represent connected (under some
definition) subsets of a network. We adopt the term community
instead of module in anticipation of the ``igraph'' package
\cite{igraph} in the ``R'' software system \cite{R-software}
becoming the primary network analysis tool in PER.

This study explored the following research questions:
\begin{description}
\item[RQ1] Are the results of Module Analysis for Multiple-Choice
Responses replicable for large FCI data sets? If not, what changes
to the algorithm are required to detect meaningful communities of
incorrect answers? \item[RQ2] How do the communities detected
change as network-building parameters are modified? Do these
changes support the existence of a coherent non-Newtonian
conceptual model? \item[RQ3] How is the incorrect answer community
structure different between the pretest and the post-test?
\item[RQ4] How is the incorrect answer community structure
different for men and women? Do the differences explain the gender
unfairness identified in the instrument?
\end{description}

This work extends the module analysis technique to a larger data
set, explores alternate choices during that analysis, and
contrasts structure between pre- and post-test data. Structural
clues in the community structure are examined to explain
unresolved questions about gender differences in answer choices
\citep{traxler2018}.

\section{Methods}

\subsection{Instrument}

The FCI is a 30-item instrument designed to measure a student's
facility with Newtonian mechanics \cite{hestenes1992}. The
instrument includes items involving Newton's three laws as well as
items probing an understanding of one- and two-dimensional
kinematics. The instrument was also constructed with distractors
representing common student misconceptions.

\subsection{Sample}

\label{sec:samples} The data for this study was collected at a
large southern land-grant university serving approximately 25,000
students. Overall university undergraduate demographics were 79\%
White, 5\% African American, 6\% Hispanic, and other groups each
with 3\% or less \cite{usnews}.

The sample was collected in the introductory calculus-based
mechanics class serving primarily physical scientists and
engineers. The sample has been analyzed previously by Traxler {\it
et al.} (Study 2, Sec. \ref{sec:study2}) \cite{traxler2018}; it is
referenced as Sample 1 in that work. The sample contains 4716
complete FCI post-test records (3628 men and 1088 women) and 4509
complete pretest records (3482 men and 1027 women). Table II in
Study 2 reports basic descriptive statistics. On the pretest, men
have an average percentage score of 43\%, women 32\%. On the
post-test, men have an average percentage score of 73\%, women
65\%. The course in which the sample was collected was presented
using the same pedagogy and managed by the same lead instructor for
the period studied. A more thorough discussion of the sample and
the instructional environment may be found in Study 2.

\subsection{Analysis Methods}

Initial replication of Study 1 was performed with the Infomap
software available from mapequation.org \cite{mapequation}.
All other statistical analysis was performed in the ``R'' statistical
software system \cite{R-software}. This work failed to replicate
the Study 1 results and proposes a modified analysis method; as
such, the analysis method is a result of the work and the various
network techniques employed are described as they are used. 

\section{Results}

\begin{figure*}[h]
    \centering
    \includegraphics[width=0.9\textwidth]{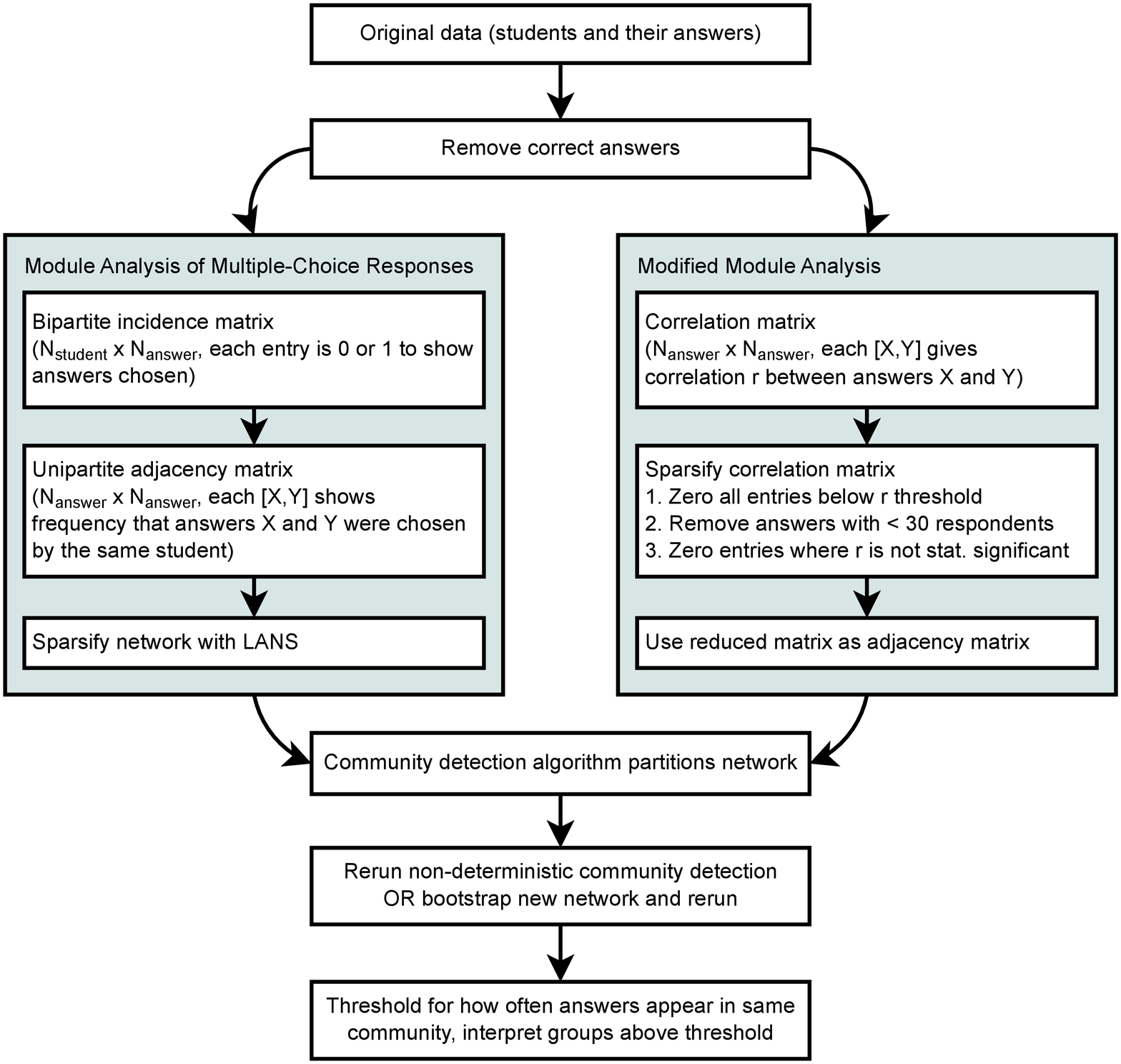}
    \caption{Workflow of analysis for the original module analysis method (left branch) and our modified version (right branch).\label{fig:analysis}}
\end{figure*}

\label{sec:ModuleAn}
\subsection{Module Analysis}
\begin{figure}[h]
    \centering
    \includegraphics[width=3.5in]{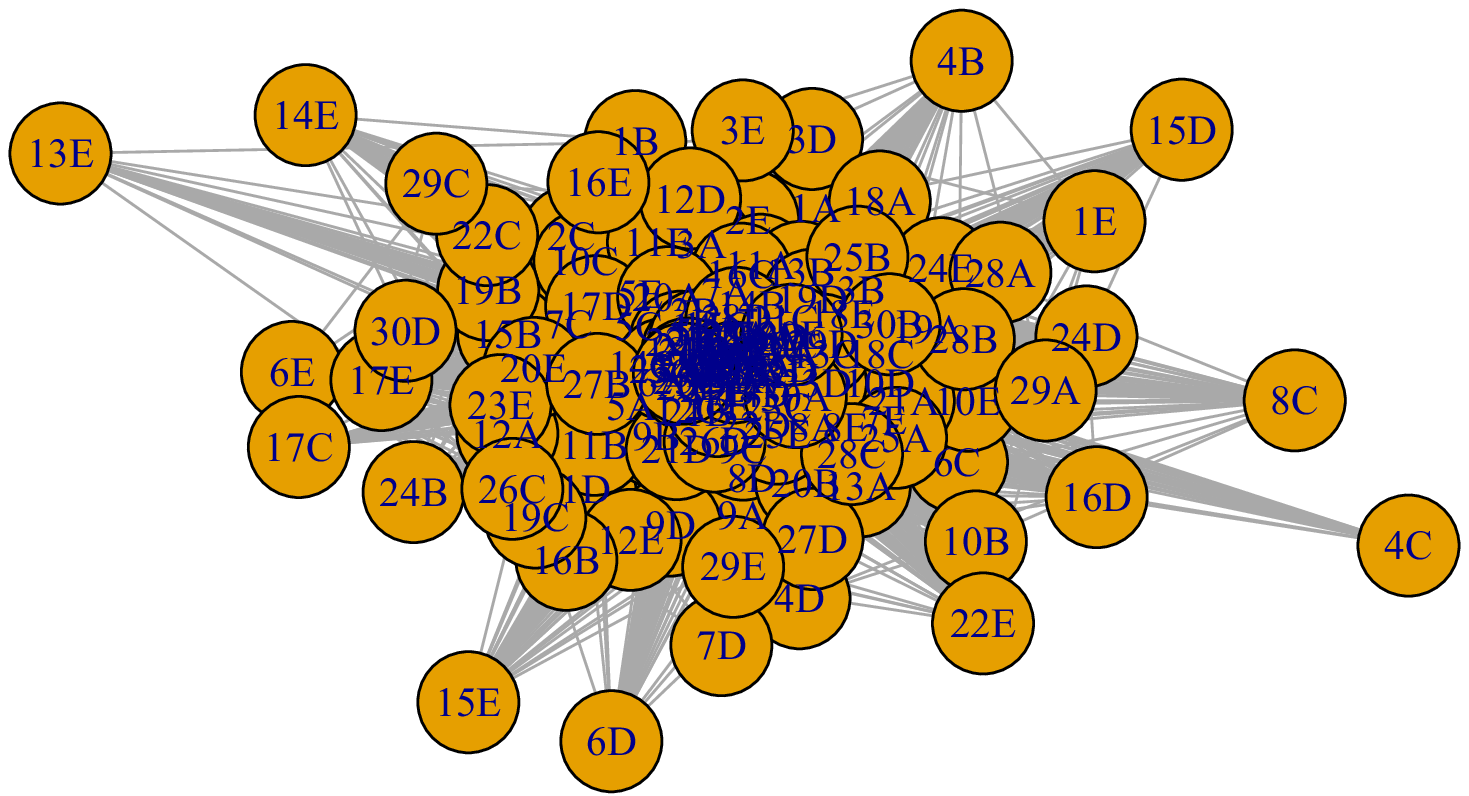}
    \caption{Unipartite network for the FCI post-test responses of women.\label{fig:adjmat}}
\end{figure}

Figure \ref{fig:analysis} outlines the original and modified
analysis steps. The original module analysis method presented in
Study 1 first formed a bipartite network, a network that includes
two types of nodes where all edges connect nodes of different
types. This network included nodes representing students and nodes
representing FCI responses. The bipartite network is then
projected into a unipartite network containing only nodes
representing FCI responses. Edges in this network connect
different responses of the same student. Edge weights represent
the number of students who selected the pair of responses
connected by the edge. For example, if $40$ students selected FCI
responses 1A and 2B, where the number is the item number and the
letter is the response within the item, there would be an edge
between node 1A and node 2B with weight $40$. While the bipartite
network can be used to extract additional properties of the
network \cite{borgatti2011analyzing}, this was not done in Study
1. As such, we began with the unipartite network. The unipartite
network can be represented by a two-dimensional matrix, called the
adjacency matrix, $adj(X,Y)$, where $X$ and $Y$ are FCI item
responses (for example, $X=1$A). The value $adj(X,Y)$ is the
number of students who selected response $X$ and response $Y$. In
the above example, $adj(1A, 2B)=40$. The network representing the
post-test responses of women on the FCI post-test is shown in Fig.
\ref{fig:adjmat}.   Because of the differences identified in men
and women in Study 2 for this sample, all results are reported
disaggregated by gender. The network in Fig. \ref{fig:adjmat} is
fairly representative of the pretest and post-test networks for
men and women. Figure \ref{fig:adjmat} uses a node placement
algorithm that places more densely connected nodes close to one
another. As in Study 1, only incorrect responses were included in
the network. The correct responses are highly correlated and are
often the most commonly selected responses. If they are included
in the network, they form a tightly connected community that
prevents exploration of the incorrect answers.

To attempt to replicate the results of Study 1, community
detection algorithms were applied to the network shown in Fig.
\ref{fig:adjmat}. First, a complete replication was attempted
which employed the ``Infomap'' software available at
``mapequation.org'' \cite{mapequation} as was originally used in
Study 1. This software, designed for very large networks, presents
such significant installation and use barriers that it seems
unlikely that it will ever achieve broad acceptance in PER. A
second path to replication using the ``infomap'' implementation in
the ``igraph'' package \cite{igraph} in ``R'' was also attempted.

To extract meaningful structure from a high-density network, the network must
be generally be simplified without removing important structure.
The process of simplifying a network by removing edges is called
``sparsification.'' The network sparsification method used in
Study 1 was Locally Adaptive Network Sparsification (LANS)
\cite{foti}. The LANS algorithm removes edges based on the
distribution of edge weights connected to each node. The
probability of selecting an edge with a smaller weight at random
is compared to a predetermined significance level
and only edges above that level are retained.  This method is
locally adaptive because it depends only on the edges incident on
a single node. A consequence of sparsifying based on the
distribution of weights incident on each node is that no node will
have its last edge removed, so no connected node is unconnected
from the rest of the network. This ensures that local structures
important to the global structure of the network are retained. 

After sparsifying with LANS (using code from
\citet{traxler_networks_2018}, Supplemental Material), the Infomap
CDA was applied. Infomap is based on information theoretic
methods. Infomap records a random walk through the network by
assigning codewords to each node, then trying to minimizing the
length of the description. Nodes visited more often are given
shorter codes and coherent communities, where the random walker
tends to spend more time, are given their own unique codes to
further reduce the information needed to represent the network.
This way, the codes for individual nodes can be reused within the
larger community structure. Nodes in one of the large communities
are connected more to each other than to nodes outside the
community. Because Infomap is not deterministic, it was run 1000
times and the communities that were most often found were selected
as the misconception modules in Study 1.

Applying Infomap with LANS sparsification failed to identify
meaningful community structure for the large data set in the current
study; Infomap consistently identified only one large community.

To explore the source of the discrepancy with Study 1, an
alternate implementation of Infomap was employed; this
implementation was part of ``igraph'' package in the ``R''
software system. A simpler sparsification algorithm was also
employed. The LANS algorithm statistically evaluates each edge,
but will not remove the last edge connecting a node. This
algorithm is a reasonable choice for a network where every edge is
purposeful (such as air travel), but may amplify noise in a
network of student responses where some edges are the results of
careless mistakes or guessing. As such, the network was also
sparsified by imposing a threshold requiring edges to have a
minimum weight. Multiple thresholds were tried. The Infomap
community detection algorithm used in Study 1 identified only one
community at all threshold values. Many other CDAs are available
in the ``igraph'' package; some identified two communities even at
very high thresholds. No CDA identified more than 2 communities.

\begin{figure}[h]
    \centering
    \includegraphics[width=3.5in]{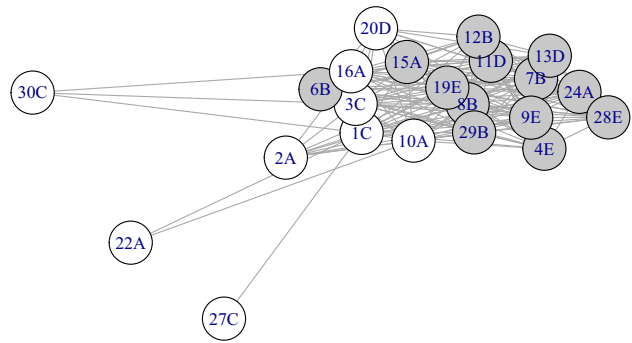}
    \caption{Communities detected for the adjacency matrix of FCI post-test responses of women
    with an edge weight threshold of $N/2$ using the fast-greedy CDA.
    \label{fig:adjmatcom}}
\end{figure}

Fig. \ref{fig:adjmatcom} shows the communities identified by the
``fast greedy'' CDA at an edge weight threshold of $N/2$ where $N$
is the number of participants; only $22$ nodes remain connected at
this threshold. Nodes in different communities are shown in
different colors. The fast-greedy CDA \cite{newman:2004b} is an
improved version of a modularity-based CDA algorithm
\cite{newman:2004a}. Modularity is a measure that compares, for a
given division of a network into communities, how many more
intra-community links exist than expected by chance in an
equivalent network \cite{newmangirvin:2004}. Modularity values
range from zero to one, where a network with a modularity of zero
means there is no clustering in the network and a modularity of
one is a strongly clustered network.

There seem to be two likely sources of the differences of the results of
this study and Study 1: sample size and the LANS algorithm. To
investigate sample size, 100 subsamples of 143 students each were
drawn from the sample in this study. Applying Infomap using ``R''
identified only one community 100\% of the time with no
sparsification and one community 92\% of the time with the
requirement the edge weight be at least N/10 where $N$ is the
number of students.

The igraph package implements many CDA algorithms; for the small
network analyzed in this work, most perform similarly. For rest of
this work, the ``fast-greedy'' CDA described above will be used.
Again, the data was subsampled to 143 students to compare with
Study 1. With no sparsification, the fast-greedy algorithm
identified 3 to 6 communities with 3 to 4 communities identified
in 92\% of the runs. With the edge weight greater than $N/10$
sparsification, fast-greedy identified 2 to 4 communities with
66\% of the runs identifying 3 communities. The communities
identified made little theoretical sense within the framework of
Study 3 with very different items in the same communities. As
such, while some of the differences in the studies may be
attributed to sample size, the choice of CDA also influenced the
communities identified at small sample size. At the large sample
size of the current study, the various community detection
algorithms implemented in igraph give fairly similar results.

\subsection{Correlation Analysis}
Part of the cause of the failure of MAMCR to find meaningful
community structure for large samples can be understood by
comparing the adjacency matrix to the correlation matrix. The correlation
matrix also defines a network, most usefully when a threshold value is applied.
The adjacency matrix which produced the network in Fig.
\ref{fig:adjmat} has no obvious clustered structure. The partial
correlation matrices reported in Study 3 clearly show clustering
into distinct communities.

The correlation between item X and item Y is defined as:
\begin{equation}
\label{eq:corr} corr(X,Y)=\frac{E[(X-\mu_X)(Y-\mu_Y)]}{\sigma_X
\sigma_Y}
\end{equation}
where $\mu_j$ is the mean of variable $j$, $\sigma_j$ is the
standard deviation, and $E[Z]$ is the expectation value of the
random variable $Z$. The expectation value is defined as:
\begin{equation}
E[X]=\sum_i \frac{X_i}{N}
\end{equation}
where $i$ is a participant and $N$ is the number of participants.
Equation \ref{eq:corr} can be simplified to produce Eq.
\ref{eq:corr2}:
\begin{equation}
\label{eq:corr2} corr(X,Y)=\frac{E[X\cdot Y] - \mu_X\cdot
\mu_Y}{\sigma_X \sigma_Y}
\end{equation}

For dichotomously scored items, the sum $\sum_i X_i Y_i$ is the
X,Y entry in the adjacency matrix, $adj(X,Y)=\sum_i X_i Y_i$. The
correlation matrix is then related to the adjacency matrix by the
expression:
\begin{equation} corr(X,Y)=\frac{adj(X,Y) - N
\mu_X\cdot \mu_Y}{N\sigma_X \sigma_Y}
\end{equation}

A pair of items can have a large $adj(X,Y)$ in a number of ways:
(a) purposeful association, students preferentially select the two
items together, or (b) accidental association, many students
select both items so on average the items get selected together
often. By subtracting the product of the means, the correlation
matrix eliminates the second case and only has large values for
purposefully selected pairs. This suggests the adjacency matrix
contains many more edges that are the result of random chance than
the correlation matrix. The correlation matrix also has the
substantial advantage of the existence of significance tests for
entries allowing the discarding of non-significant edges.

With this observation, we propose a modification of MAMCR, called
Modified Module Analysis (MMA), that investigates the community
structure of the correlation matrix. The remainder of the this
work investigates this proposal. The differences between MAMCR and
MMA are presented schematically in Fig.
\ref{fig:analysis}. 

\begin{figure*}[h]
    \centering
    \includegraphics[width=6.3in]{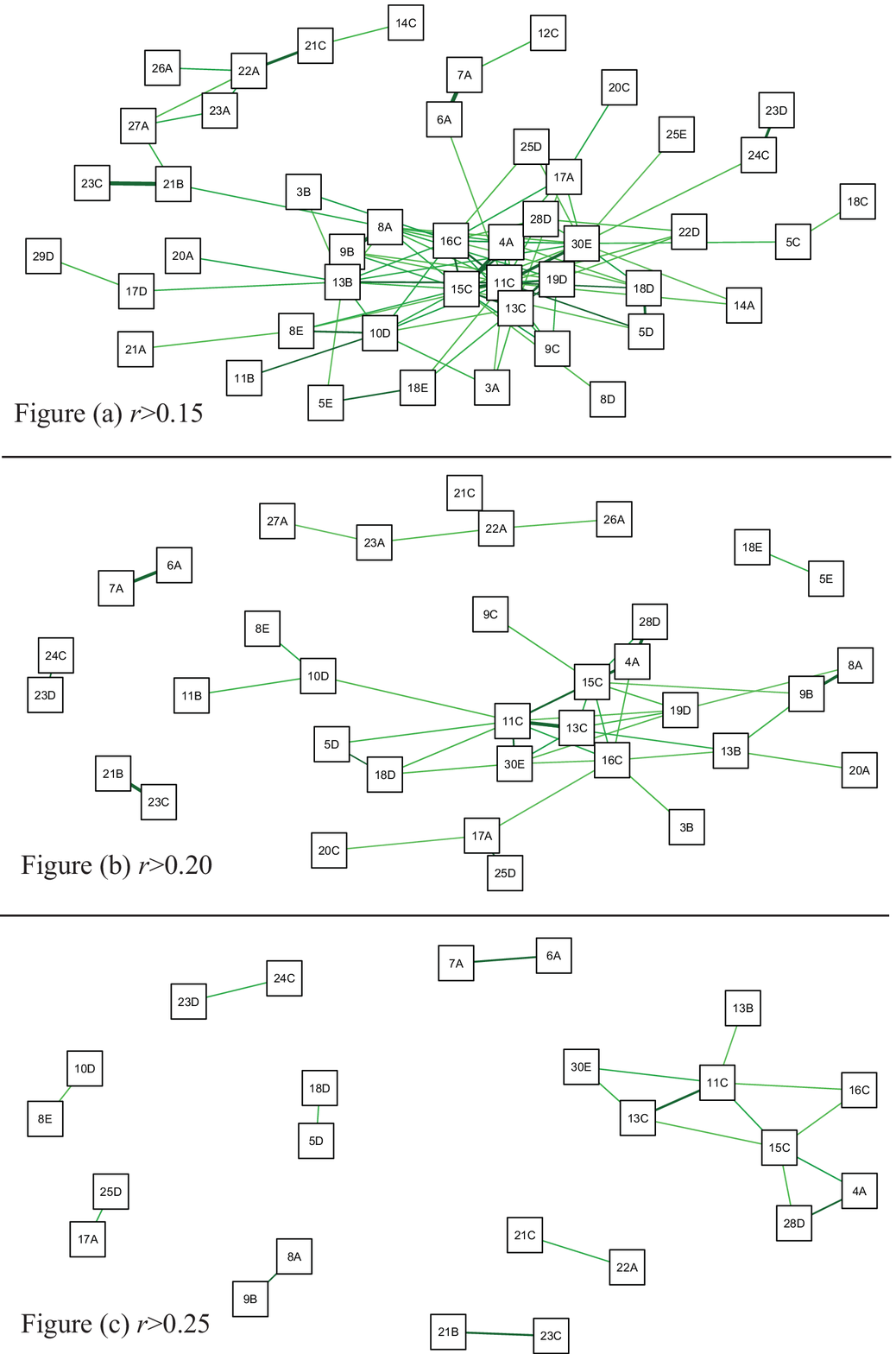}
    \caption{\label{fig:corplots} Post-test correlation matrices of women at varying levels of $r$.}
\end{figure*}

To explore this proposal, the correlation matrix was calculated
for all incorrect answers. Nodes with too few participants to be
statistically reliable were eliminated; for this work, nodes with
fewer than $30$ responses were removed. Edges were removed where
the correlation, $r$, between the two nodes was not significant at
the $p=0.05$ level where a Bonferroni correction was applied to
reduce the Type I error rate. As with the adjacency matrix, a
threshold was then applied to simplify the network. For this work,
only positive correlations were considered; future work will
investigate bipartite networks with positive and negative
correlations. Figure \ref{fig:corplots} shows the correlation
matrix for the post-test results of women retaining only entries
with $r>0.15$, $r>0.20$, and $r>0.25$. The representation in Fig.
\ref{fig:corplots} was produced by the ``qgraph'' package in ``R''
\cite{qgraph}. The width of the line is proportional to the size
of the correlation. Node placement is for visual effect only.

\subsection{Modified Module Analysis}

\begin{figure}[h]
    \centering
    \includegraphics[width=3.5in]{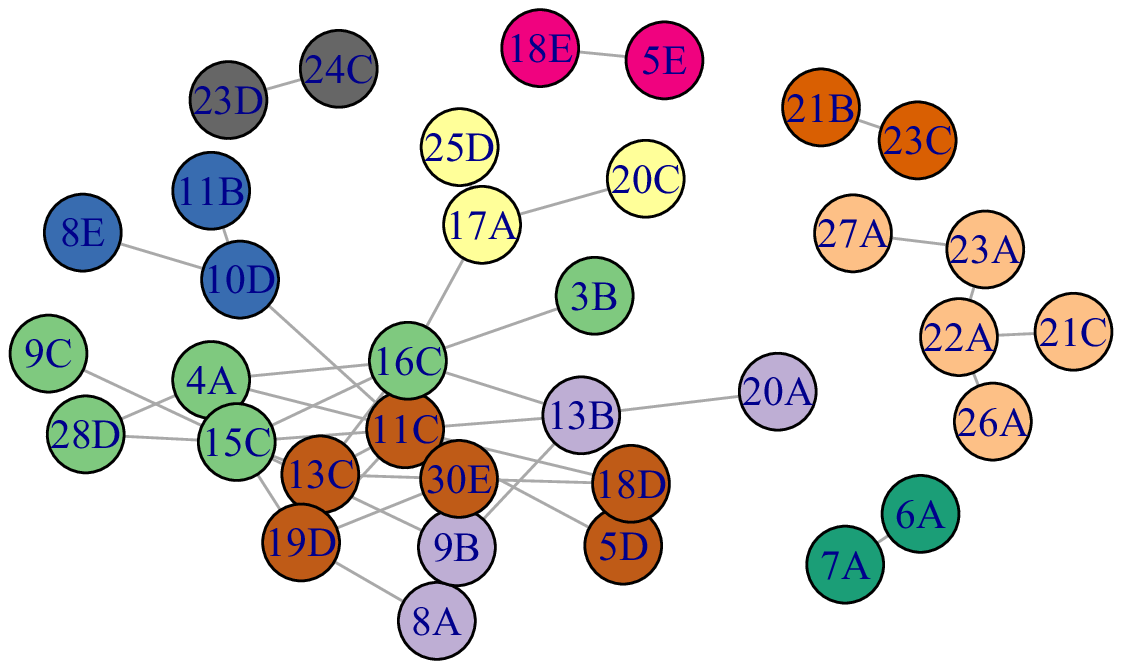}
    \caption{Communities detected in the FCI correlation matrix with $r>0.2$. Each community is drawn in a different color.
    \label{fig:cor-com}}
\end{figure}

The correlation matrices in Fig. \ref{fig:corplots} show a clear
clustered structure. MMA was applied to understand these
structures. The communities detected for the $r>0.20$ correlation
matrix are shown in Fig. \ref{fig:cor-com}. Figure
\ref{fig:cor-com} shows the communities identified by a single
application of the fast-greedy CDA. To understand the stability of
these structures, the algorithm was applied multiple times. The
communities identified by the multiple applications of the
algorithm presented later in the paper do not fully align with those in Fig.
\ref{fig:cor-com}.

Both the CDA and the sample itself contain randomness, and
therefore, some of the community structure in Fig.
\ref{fig:cor-com} may result from chance. To determine the part of
the community structure not resulting from random fluctuations,
bootstrapping with 1000 replications was performed. Bootstrapping
is a statistical technique that forms a distribution of a
statistic of interest by subsampling the data set with replacement
\cite{boot2}. The ``R'' ``boot'' package was used to perform the
bootstrapping \cite{boot1}. Part of the goal of this work was to
compare the community structure of men and women. As one part of
the sparsification process, infrequently-selected nodes and
insignificantly weighted edges were removed. Both the threshold
for an infrequently-selected node and edge significance depend on
sample size. For a fair comparison, a sample balanced between men
and women was required. The full data set was very unbalanced. To
correct for this, when the male sample was bootstrapped, 1000
samples were drawn each of the same size as the overall female
sample. For the female sample, bootstrapping was performed by
subsampling with replacement which preserved the overall size of
the sample.

The number of times each pair of responses was found in the same
community was recorded for each of the 1000 samples forming a
``community matrix.'' The community matrix was nearly completely
disconnected into small clusters. The clusters are shown in Table
\ref{tab:commat}. Responses were identified in the same
communities at different rates. We define the community fraction,
$C$, as the fraction of the bootstrap subsamples in which the pair
of items were found in the same community. The community matrix
was filtered to show items that were identified in $C>60\%$ and
$C>80\%$ of the communities in the 1000 bootstrap replications in
Table \ref{tab:commat}.
The majority of the communities extracted from the community
matrix were fully connected; each node was connected to every
other node in the community. Some, however, were not. The
intra-community density, $\gamma$, is defined as the ratio of the
number of edges in the community to the maximum number of edges
possible \cite{zweig}. For communities with $\gamma<1$, $\gamma$
is presented as a percentage in parenthesis in Table
\ref{tab:commat}. For example, if a community contains four nodes
then there are a maximum of six distinct edges between the nodes.
If the community only possesses five of those edges, then
$\gamma=5/6$.

\begin{table}[h]
        \caption{Communities identified in the pretest and post-test incorrect answers at $r>0.2$ and differing levels of the community fraction, $C$. The number in parenthesis is the
        intra-community density, $\gamma$,
        for communities where the intra-community density is not one. \label{tab:commat} }
    \centering
    \begin{tabular}{|l|cc|cc|}
        \hline
\multirow{2}{*}{Community}&\multicolumn{2}{|c}{Pretest}&\multicolumn{2}{|c|}{Post-test}\\
&Men&Women&Men&Women\\\hline
\multicolumn{5}{|c|}{$C>60\%$}\\\hline
1A,2C,15B,19B&X(67\%)&&&\\
1D,2D&X&X&&\\
3B,13B&&&X&\\
4A,15C,28D&&X&X&X\\
4A,15C,16C,28D&X&&&\\
5D,11C,13C,18D,30E&&&X&\\
5D,11C,13C,18D,19D,30E&&&&X(73\%)\\
5E,18E&X&X&X&X\\
6A,7A&X&X&X&X\\
8A,9B&X&X&X&X\\
8E,10D&&&&X\\
11B,29A&X&X&&\\
15D,16D&X&X&&\\
17A,25D&&&X&X\\
21B,23C&&&X&X\\
21C,22A,23A,26A&&&&X(83\%)\\
21C,22A&X&&X&\\
23D,24C&X&X&X&X\\\hline \multicolumn{5}{|c|}{$C>80\%$}\\\hline
1A,2C&X&&           &\\
1D,2D&X&&           &\\
4A,15C,28D&X&X&     X&X\\
5D,11C,13C,18D,30E&&&&X(60\%)\\
5D,18D&&&            X&\\
5E,18E&X&X&         X&X\\
6A,7A&X&X&          X&X\\
8A,9B&X&X&          X&X\\
11B,29A&X&X&        &\\
11C,13C,30E&&&       X&\\
17A,25D&&&           X&X\\
21B,23C&&&           X&X\\
21C,22A&X&&         X&X\\
23D,24C&X&X&         X&X\\\hline
    \end{tabular}
\end{table}

\subsection{The Structure of Incorrect FCI Responses}

Unless otherwise stated, results below are reported for $C>0.8$ and $r>0.2$.

\subsubsection{Types of Incorrect Communities}

\def\addmc{{\bf (Add) }}
\def\removemc{{\bf (Remove) }}

\begin{turnpage}
\begin{table*}[h]
        \caption{Misconceptions represented by incorrect answer communities. Communities marked with a $*$ result from blocked problems. Proposed additions are marked (Add). Proposed items to be
        removed are marked (Remove). If Add or Remove is placed before all items, it applies to all items. If Add or Remove is placed before only one of many items, it applies to that item.
        \label{tab:misconceptions} }
    \centering
    \begin{tabular}{|c|c|c|c|}
        \hline
\multirow{2}{*}{Community}&\multicolumn{2}{c|}{Naive Conceptions}&\multirow{2}{*}{Dominant Misconception}\\\cline{2-3}
&Category&Sub-Catagory&\\\hline
\multirow{2}{*}{1A, 2C}&\multirow{2}{*}{Gravity}&1A (G3): Heavier objects fall faster&\multirow{2}{*}{Heavier objects fall faster} \\
&&\addmc 2C:  Heavier objects travel farther&\\\hline
\multirow{2}{*}{1D, 2D}&\multirow{2}{*}{ Gravity}&\addmc 1D: Lighter objects fall faster&\multirow{2}{*}{Lighter objects fall faster}\\
&&\addmc 2D: Lighter objects travel farther&\\\hline
\multirow{2}{*}{4A, 15C, 28D}&\multirow{2}{*}{Action/Reaction Pairs}& 4A, 28D (AR1): Greater mass implies greater force &Greater mass implies greater force \\
&&15C, 28D (AR2): Most active agent produces greatest force&Most active agent produces greatest force\\\hline
\multirow{3}{*}{5D, 18D}&\multirow{3}{*}{\shortstack{Impetus\\ Active Forces}} &\removemc 5D, 18D (I5): Circular impetus &\multirow{3}{*}{Motion implies active forces} \\
&&5D (I1): Impetus supplied by ``hit''&\\
&& 5D, \addmc 18D (AF2): Motion implies active forces&\\\hline
\multirow{4}{*}{5E, 18E}&\multirow{4}{*}{\shortstack{Impetus\\ Active Forces}} &5E (I1): Impetus supplied by ``hit"&\multirow{4}{*}{\shortstack{Motion implies active forces\\ Centrifugal force}} \\
&&\removemc 5E (I5): Circular impetus&\\
&&5E, \addmc 18E (I1): Motion implies active forces&\\
&&5E, 18E (CF): Centrifugal force&\\\hline
6A,7A&Impetus& 6A, 7A (I5): Circular impetus& Circular impetus\\\hline
8A, 9B$*$&Concatenation of Influences& 8A, 9B (CI3): Last force to act determines motion&\\\hline
\multirow{3}{*}{11B, 29A}&\multirow{3}{*}{\shortstack{Other Influences on Motion\\ Impetus}} & 29A, \removemc 11B (Ob): Obstacles exert no force&\multirow{3}{*}{Motion implies active forces}\\
&&\removemc 11B (I1): Impetus supplied by ``hit"&\\
&&\addmc 11B (AF2): Motion implies active forces& \\\hline
\multirow{3}{*}{11C, 13C, 30E}&\multirow{3}{*}{Impetus}&11C, 30E (I1): Impetus supplied by ``hit'' & \multirow{3}{*}{Motion implies active forces}\\
&&\addmc 11C, 13C (AF2): Motion implies active force&\\
&&13C (I3): Impetus dissipation&\\\hline
\multirow{2}{*}{17A, 25D}&Concatenation of Influences & 17A, \addmc 25D (CI1): Largest force determines motion  &\multirow{2}{*}{Largest force determines motion}\\
&Resistance&25D (R2): Motion when force overcomes resistance&\\\hline
21B, 23C*&Concatenation of Influences& 21B, 23C (CI3): Last force to act determines motion&\\\hline
\multirow{2}{*}{21C, 22A*}&Concatenation of Influences & 21C (CI2): Force compromise determines motion &\\
&Active Forces&22A (AF4): Velocity proportional to applied force&\\\hline
\multirow{2}{*}{23D, 24C*}&\multirow{2}{*}{Impetus}&23D, 24C (I3): Impetus dissipation& \multirow{2}{*}{Impetus dissipation}\\
&& 23D (I2): Loss/recovery of original impetus&\\\hline
    \end{tabular}
\end{table*}
\end{turnpage}

Table \ref{tab:misconceptions} classifies the incorrect reasoning
for each community of incorrect answers 
in Table \ref{tab:commat}. These can be divided into two general
classes: communities resulting from blocking and communities
resulting for consistently applied incorrect reasoning
(misconceptions). Communities \{8A, 9B\}, \{21B, 23C\}, and \{21C,
22A\} are answers within blocked problems where the second answer
in the pair would be correct if the first answer was correct. The
other communities apply either the same incorrect reasoning or
related incorrect reasoning.

Hestenes and Jackson produced a detailed taxonomy of the naive
conceptions (their terminology) tested by the FCI  \cite{fci-table}. Table
\ref{tab:misconceptions} shows a mapping of this taxonomy onto the
incorrect answer communities identified in the current work. The
taxonomy divides the naive conceptions into a general category and
a number of sub-categories. The number in parenthesis in Table
\ref{tab:misconceptions} is the sub-category label
\cite{fci-table}. Items marked with an asterisk in Table
\ref{tab:misconceptions} are part of item blocks. Because the
relation between the items seems to be largely generated by the
interdependencies resulting from blocking rather than consistently
applied misconceptions, the blocked items will not be discussed
further.

Some issues arise in comparing the proposed FCI taxonomy with
communities identified by MMA and the similar item blocks
identified in Study 3. First, for some of items in the incorrect
communities, no misconception was identified (items 1D, 2C, and
2D). Students are answering these items in a correlated manner
which implies the possibility of consistent reasoning patterns;
for these items a possible misconception was suggested. The new
misconception was labeled ``(Add).''
Table \ref{tab:misconceptions} shows that often the items in the
incorrect communities identified by MMA belong to multiple naive
misconception categories and have different sub-categories. This
would seem to imply that the naive conception taxonomy is more
detailed than the actual application of misconceptions by students
as measured by the FCI. For example, in the Newton's 3rd law
community, \{4C, 15C, 28D\}, different items involve objects of
different activity from one student pushing on another student
(item 28, one active object), to a car pushing a truck (item 15,
one active object), to a head-on collision (item 4, two active
objects). This distinction does not seem important to the
student's answering pattern. It is unclear if this results from
one misconception taking precedence over another or from students
applying more general reasoning as proposed by the resource or
knowledge-in-pieces models. As such, Table \ref{tab:misconceptions} includes a
column which proposes a title for the dominant misconception. In many cases, the dominant misconception was identified as
the misconception shared by the majority of the items. In some cases, a
dominant misconception was proposed. For the Newton's 3rd law
community, \{4C, 15C, 28D\}, multiple misconceptions were shared equally and no
dominant misconception was identified. In the following, the combination of the
greater mass implies greater force and most active agent produces greater force misconceptions
are called ``Newton's 3rd law misconceptions.''

The coding of the misconceptions represented by the 5D and 18D
community seems problematic. Item 5 represents a ball shot into a
circular channel and item 18 a boy swinging on a rope. Answer D on
both items includes a force in the direction of motion, and
therefore, it is unclear why item 18 is not included in the motion
implies active forces misconception. This
has been added to Table \ref{tab:misconceptions}.

Items 17 and 25 also require some additional analysis. Both
involve objects moving at a constant speed under the influence of
multiple forces. In both 17A and 25D, the greater force is in the
direction of motion. It seems that 25D should also test the
largest force determines motion misconception.

Multiple items were identified as involving the misconception of
circular impetus. Circular impetus is used in two alternate ways.
In responses 6A and 7A, circular impetus involves an object
continuing to move in a circle after a constraint is removed. In
responses 5D, 5E, 18D, and 18E, the constraint is still in place.
Both items 5 and 18 also include a force of the channel or rope in
the list of forces; these forces are unnecessary if the object
moves in a circle of its own accord. As such, we propose removing
the circular impetus misconceptions from 5D, 18D, 5E, and 18E; these items
have been labeled (Remove) in Table \ref{tab:misconceptions}.
Item 18 also does not provide a response that includes both the
centrifugal force and the motion implies active forces
misconceptions; as such, students may still be applying the motion
implies active force misconception, it is just not tested by item
18E.

The pretest community \{11B, 29A\} is also curious. In item 11, a
hockey puck is struck activating the impetus supplied by the
``hit'' misconception, but response 11B explicitly asks about a
force in the direction of motion. As such, we propose this item
also tests the motion implies active forces misconception. It is also unclear
how item 11B probes the obstacle exerts no force misconception; we propose it
be removed from the item. Item 29
involves a chair sitting on a floor; response 29A identifies only
the force of gravity on the object and ignores the normal force.
It seems difficult to claim this community probes a common
misconception. Item 29 was also demonstrated to have poor
psychometric properties in Study 2; the correlation between 11B
and 29A may have resulted from 29A not functioning as intended.

The community \{11C, 13C, 30E\} continues to convolve the motion
involves active forces misconception with the impetus supplied by
the ``hit'' misconception. Response 30E explicitly discusses the
force of the ``hit'' while items 11C and 13C discuss a force in
the direction of motion. We propose adding this misconception to
the items 11C and 13C. Further, only item 13C involves the idea of a
dissipation of impetus. For this community, while multiple
misconceptions are tested, one seems to dominate student
responses, motion implies active forces.

Finally, the blocked item responses 23D, 24C differ from the other
blocked responses. Rather than the second response being the
correct answer if the first response was correct, both appear to
be applications of the dissipation of impetus misconception.

\subsubsection{Reducing Sparsification}

The $r>0.2$ and $C>0.8$ thresholds generated a fairly disconnected
network. This network was productive in identifying incorrect
answers that were frequently selected at the same time by the same
student. As these thresholds are relaxed, the network becomes more
connected as shown in Fig. \ref{fig:cor-com}. As the network
becomes more connected, related misconceptions may merge showing
the students have a coherent non-Newtonian force concept. The
communities identified at
$r>0.15$ and
for $C>0.6$ and $C>0.8$ are shown in the Supplemental Material \cite{supp}.
While relaxing
the thresholds did allow the community \{5D, 11C, 13C, 18D, 30E\}
to be detected for both men and women, most other new communities
identified did not result from the merger of communities
identified at more restrictive thresholds. Particularly on the
pretest, the larger communities do not make much sense in terms of
the framework of Study 3. This is particularly evident in the
mixing of the Newton's 3rd law items \{4, 15, 16, and 28\} with
other items. As such, it appears that student misconceptions exist
relatively independently as small groups of consistent answers,
not as a part of a larger coherent framework.

\begin{table*}[t]
        \caption{\label{tab:misc} Percentage of students selecting each incorrect community for the FCI post-test. A t-test was performed to
        determine if the differences between men and women were significant, the $p$-value is presented.
        Cohen's $d$ for the difference is also presented. }
    \centering
    \begin{tabular}{|l|cc|c|c|c|}\hline
    \multirow{2}{*}{Community}&Male&Female&\multirow{2}{*}{$p$}&\multirow{2}{*}{$d$}&\multirow{2}{*}{Misconception}\\
    &Ave. (\%)&Ave. (\%)&&&\\\hline
    4A, 15C, 28D& $32\pm 47$& $33\pm 47$&0.27&0.02&Newton's 3rd law misconceptions\\
    5D, 11C, 13C, 18D, 30E& $22\pm 42$& $20\pm 40$&$<0.001$&0.06&Motion implies active forces\\
    5E, 18E& $7\pm 25$& $7\pm 25$&0.69&0.01&Motion implies active forces, centrifugal force\\
    6A, 7A& $14\pm 35$& $5\pm 21$&$<0.001$&0.39&Circular impetus\\
    17A, 25D& $42\pm 49$& $37\pm 48$&$<0.001$&0.11&Largest force determines motion\\\hline
    \end{tabular}
\end{table*}

\subsubsection{The Strength of Common Misconceptions}

One motivation of Study 1 was to provide instructors with a
mechanism for identifying common misconceptions so that specific
interventions could be targeted to address those misconceptions.
The communities of incorrect answers remaining on the post-test as
shown in Table \ref{tab:commat} could be used provide a measure of
the prevalence of the misconception in the classes studied. Table
\ref{tab:misc} presents an overall average for each incorrect
community in Table \ref{tab:commat} on the post-test. Only
communities that did not result from problem blocking are
presented. Averages were calculated by assigning a score of $1$ if
the response was selected and $0$ if it was not, then averaging
over each item in the group. Results are disaggregated by gender
and the $p$-value for a $t$-test to determine if differences by
gender are significant is also presented; Cohen's $d$ provides a
measure of effect size. Cohen suggests $d=0.2$ as a small effect,
$d=0.5$ as a medium effect, and $d=0.8$ as a large effect
\cite{cohen}.

The overall difference in post-test percentage score between men
and women was 8\%. The percentage of students who answer an item
correctly directly influences the percentage of students who
answer an item incorrectly; therefore, only differences in Table
\ref{tab:misc} greater than 8\% represent unexpected differences
between men and women. Only items \{6A, 7A\} exceed this
difference, but then only slightly with a difference of 9\%. Items
\{6A, 7A\} are also the only community with differences of at
least a small effect size; however, the effect size is likely
inflated by the small standard deviation of women because of a
floor effect. In general, the rate of selecting one of the
communities of common incorrect answers was very similar for men
and women.

For the class studied, the results of Table \ref{tab:misc} suggest
that additional effort be directed to addressing the largest force determines motion misconception
measured by \{17A, 25D\} and Newton's 3rd law misconceptions measured by \{4A, 15C, 28D\}.

\section{Discussion}
\subsection{Research Questions}
This study sought to answer four research questions; they will be
addressed in the order proposed.

{\it RQ1: Are the results of Module Analysis for Multiple-Choice
Responses replicable for large FCI data sets? If not, what changes
to the algorithm are required to detect meaningful communities of
incorrect answers?} The MAMCR process described in Study 1
identified only one or two communities in our data whether using
LANS or an edge weight threshold to sparsify the network. This
result held for Infomap and for other CDAs. Reducing the data to a
comparable size by subsampling generated more communities, but still fewer than
identified in Study 1; however, the communities identified did not make
conceptual sense. We concluded that the community structure
identified in Study 1 was the result of the low sample size and
the LANS algorithm and that modifications to MAMCR were needed to
productively identify incorrect answer communities.

The failure of MAMCR for large samples led us to propose a variant
of the algorithm using the correlation matrix instead of the
adjacency matrix to build the network. This matrix was sparsified
by removing statistically insignificant correlations and
correlation below a threshold ($r<0.2$ for most of our analyses).
Using the fast-greedy CDA on this new network, produced a rich set
of incorrect communities (Table \ref{tab:commat}). These
communities were often related to items with related correct
answers as identified by Study 3. These communities fall into two
broad categories: those employing similar incorrect reasoning  and
those resulting from  ``blocked'' items where an incorrect choice
later in the block is the correct answer given an incorrect choice
earlier in the block.

{\it RQ2: How do the communities detected change as
network-building parameters are modified? Do these changes support
the existence of a coherent non-Newtonian conceptual model?} A
more permissive threshold for the correlation matrix ($r>0.15$)
yielded larger communities as shown in the Supplemental Material
\cite{supp}. These larger communities were not formed by the
joining of smaller communities related to the same misconception;
in fact, many of the communities contained items that had little
conceptual relation. As such, it appears that the best model of
student misconceptions are as isolated pieces of reasoning
associated with items with a similar correct solution structure.

{\it RQ3: How is the incorrect answer community structure
different between the pretest and the post-test?} For $C>0.8$ and
$r>0.2$, a total of 14 incorrect answer communities were
identified for either men or women pre- and post-instruction; 5 of
the communities were consistently identified for both genders pre-
and post-instruction. Three of these five represent consistently
applied misconceptions: \{4A, 15C, 28D\}, Newton's 3rd law misconceptions; \{5E, 18E\},
motion implies active forces and the existence of a
centrifugal force; and \{6A, 7A\}, circular impetus. The items from
which the incorrect answers in these communities were drawn were
all identified as having very similar correct solution structure
in Study 3. The other two communities were drawn from problem
blocks: \{8A, 9B\} and \{23D, 24C\}. Three incorrect communities
disappeared with instruction: for all students \{11B, 29A\}, motion implies active forces;
for men only, \{1A, 2C\} heavier
objects fall faster and \{1D, 2C\} lighter
objects fall faster. Many incorrect communities were only
identified post-instruction including \{17A, 25D\} involving items
with similar solution structure as identified in Study 3.

{\it RQ4: How is the incorrect answer community structure
different for men and women? Do the differences explain the gender
unfairness identified in the instrument?} Post-instruction, using
$C>0.8$ and $r>0.2$,
 11 communities were
 identified for either men or women; 8 of these communities were
 identified for both men and women. One of the other three
 communities was only identified for women, \{5D, 11C, 13C, 18D,
 30E\}, represents the motion implies active forces
 misconception. This community was the merger of the two
 communities only identified for men \{5D, and 18D\} and \{11C,
 13C, 30E\}; the female community was also not completely
 connected, $\gamma=0.6$. As such, women may have a slightly more
 integrated motion implies active forces misconception post-instruction,
 but in general the misconception structure of men and women is
 strikingly similar post-instruction. The differences between men
 and women
 post-instruction did not involve the unfair items identified in
 Study 2 and cannot explain the unfairness of these items.

 Pre-instruction, nine communities were identified for either men or
 women; six were identified for both men and women. All of the
 communities not shared by men and women were only identified for
 men. One of these communities, \{21C, 22A\}, was the result of
 blocking and was identified for both men and women
 post-instruction. The other two communities unique to men, \{1A,
 2C\} and \{1D, 2D\}, involve the heavier objects fall
 faster and the lighter objects fall faster misconceptions. The misconception structure of men and
 women was quite similar pre-instruction, with men holding more
 consistent misconceptions.

 \subsection{Additional Observations}

The misconception communities identified by MMA were not completely consistent with
the naive conception taxonomy provided by Hestenes and Jackson for the FCI \cite{fci-table}. Often
multiple naive conceptions were associated with the same community. This may
indicate that student reasoning is better modeled by a more general
framework such as knowledge-in-pieces or ontological categories. It may also
indicate that the FCI cannot fully resolve the detailed set of misconceptions
identified in the taxonomy.

The results of this work were not consistent with recent exploratory
analyses of the FCI \cite{scott2017conceptual,scott2018central,eaton2019} which
identified a few large factors; these factors mixed very different
correct and incorrect responses. The small communities identified in the current
work, which are partially supported by the taxonomy of Hestenes and Jackson, seem to
indicate the MMA may be a more productive quantitative method to explore
misconceptions.

\section{Implications}

Not all of the communities identified in Table \ref{tab:commat}
represent misconceptions. Some represent combinations of dependent
answers. For these combinations, the second answer is correct if
the first answer was the correct answer. This suggests that,
because of the blocking of items in the FCI, a simple scoring of
the instrument with each item as correct or incorrect may
understate a student's knowledge of the material. Previous authors
have called for reevaluating the scoring of the FCI
\cite{hudson1996re}, but not because of problem blocking.

The identification of three communities of incorrect answers that
were the result of item blocking further supports the conclusions
of Study 3 that item blocking should be discontinued in future PER
instruments because it may make the instruments difficult to
interpret statistically.

The misconception communities identified in Table \ref{tab:misconceptions} allow instructors
to determine the strength of students' misconceptions as they enter a physics
class and the remaining strength after instruction,  as shown in Table \ref{tab:misc}.
This should allow instructors to adjust their classes to address misconceptions
remaining after instructoion and to direct fewer resources to addressing misconceptions
that are not present pre-instruction.

\section{Future Work}

MMA was productive in extending the understanding of the incorrect
answer structure of the FCI; it will be extended to other
conceptual instruments including the Force and Motion Conceptual
Inventory \cite{thornton1998} and the Conceptual Survey of
Electricity and Magnetism \cite{maloney2001}.

Network analysis encompasses a broad collection of powerful
analysis techniques. The analysis in this work represents the
barest beginnings of the possibilities of these techniques. Future
research may consider networks with multiple types of nodes
(possibly correct and incorrect answers or pretest and post-test
answers) or multiple types of edges (possibly negative and
positive correlations).

\section{Conclusion}

Previous results reported for Module Analysis for Multiple-Choice
Responses (MAMCR) could not be replicated for a large sample. The
failure of the algorithm at large sample size likely results from
a combination of unpurposeful edges in the adjacency matrix at
large sample sizes and properties of the LANS sparsification
algorithm. A modification of the algorithm, Modified Module
Analysis (MMA), based on the correlation matrix was productive in
identifying useful community structure. MMA identified 11
communities on the post-test and 9 on the pretest. Most of these
communities were identified both for men and women: 8 on the
post-test, 6 on the pretest. In general, the incorrect answer
community structure identified for men and women was very similar
and could not explain the gender differences previously identified
in a subset of items in the instrument. The communities identified
at high sparsification failed to merge into larger communities
addressing similar misconceptions as sparsification was reduced
suggesting that students do not have an integrated non-Newtonian
belief system, but rather isolated incorrect beliefs strongly tied
to the type of question asked.

\begin{acknowledgments}
This work was supported in part by the National Science Foundation
as part of the evaluation of improved learning for the Physics
Teacher Education Coalition, PHY-0108787.
\end{acknowledgments}

%

\end{document}